\documentclass[aps,pre,twocolumn,showpacs]{revtex4-1}
\usepackage{amsmath,graphicx,amsbsy,amssymb,epsfig,latexsym,wasysym,pifont,mathrsfs}
\usepackage{float} \newfloat{widefig}{thp}{lop}
\usepackage{comment}
\usepackage{color}
\begin{document}

\title{Random growth lattice filling model of percolation: a crossover
  from continuous to discontinuous transition}

\author{Bappaditya Roy}%
\author{S. B. Santra}%
\email{santra@iitg.ernet.in}
\affiliation{Department of Physics, Indian Institute of Technology
Guwahati, Guwahati-781039, Assam, India.}

\date{\today}
 
\begin{abstract}
  A random growth lattice filling model of percolation with touch and
  stop growth rule is developed and studied numerically on a two
  dimensional square lattice. Nucleation centers are continuously
  added one at a time to the empty sites and the clusters are grown
  from these nucleation centers with a tunable growth probability
  $g$. As the growth probability $g$ is varied from $0$ to $1$ two
  distinct regimes are found to occur. For $g\le 0.5$, the model
  exhibits continuous percolation transitions as ordinary percolation
  whereas for $g\ge 0.8$ the model exhibits discontinuous percolation
  transitions. The discontinuous transition is characterized by
  discontinuous jump in the order parameter, compact spanning cluster
  and absence of power law scaling of cluster size
  distribution. Instead of a sharp tricritical point, a tricritical
  region is found to occur for $0.5<g<0.8$ within which the values of
  the critical exponents change continuously till the crossover from
  continuous to discontinuous transition is completed.
\end{abstract}

\pacs{64.60.ah, 61.43.Bn, 05.70.Fh, 81.05.Rm}

\maketitle

\section{Introduction}
A new era in the study of percolation has started in the recent past
developing a series of new models
\cite{epj223,Saberi20151,Boccaletti20161} such as percolation on
growing networks \cite{PhysRevE.64.041902(2001)}, percolation in the
models of contagion \cite{PhysRevLett.92.218701(2004),
  PhysRevE.70.026114(2004)}, $k$-core percolation
\cite{epl73.560(2006),*PhysRevE.93.062302}, explosive percolation
\cite{Achlioptas1453,*PhysRevLett.103.045701,*Cho20131185},
percolation on interdependent networks
\cite{havlin2010,*PhysRevE.90.012803,*Radicchi2015597}, agglomerative
percolation \cite{PhysRevE.84.066111(2011),*Christensen2012},
percolation on hierarchical structures \cite{Boettcher2012}, drilling
percolation \cite{PhysRevLett.116.055701,*PhysRevE.95.010103} and
two-parameter percolation with preferential growth
\cite{Roy010101}. In these models, instead of robust second order
(continuous) transition with formal finite size scaling (FSS) as
observed in original percolation \cite{stauffer,bunde-havlin}, a
variety of new features are noted. Sometimes the transitions are
characterized as a first-order transition \cite{PhysRevE.81.030103,
  PhysRevLett.105.035701, janssen2016,Shu2017}, sometimes a crossover
from second order to first order with a tricritical point (or region)
are observed \cite{PhysRevE.70.026114(2004),PhysRevLett.106.095703,
  PhysRevE.86.061131,PhysRevE.93.052304,Roy010101}, sometimes features
of both first and second order transitions are simultaneously
exhibited in a single model
\cite{PhysRevE.81.036110,Sheinman2015,Cho2016}, sometimes second order
transitions with unusual FSS are found to appear
\cite{PhysRevLett.105.255701, Riordan2011322, Grassberger2011,
  Bastas201454}. Such knowledge not only enriches the understanding of
a variety of physical problems but also leads to creation of newer
models beside extension of the existing models for deeper understating
of the existence of such non-universal behavior.

In this article, we propose another novel model of percolation namely
``random growth lattice filling'' (RGLF) model adding nucleation
centers continuously to the lattice sites as long as a site is
available and growing clusters from these randomly implanted
nucleation centers with a constant but tunable growth probability
$g$. RGLF can be considered as a discrete version of the continuum
space filling model (SFM) \cite{Chakraborty2014} with the touch and
stop rule in the growth process as that of the growing cluster model
(GCM) \cite{Tsakiris2011,argtouchstop}. However, RGLF displays a
crossover from continuous to discontinuous transitions as the value of
$g$ is tuned continuously from $0$ to $1$ in contrast to both SFM and
GCM which display a second order continuous transition. Below we
present the model and analyze data that are obtained from extensive
numerical computations.

\section{The Model}
A Monte Carlo (MC) algorithm is developed to study percolation
transition (PT) in RGLF defined on a $2$-dimensional ($2$D) square
lattice. Initially the lattice was empty except one nucleation center
placed randomly to an empty site. In the next time step, besides
adding a new nucleation center randomly to another empty site, one
layer of perimeter sites of all the existing active clusters including
the nucleation center implanted in the previous time step are occupied
with a constant growth probability $g$ following the Leath algorithm
\cite{leath}. The process is then repeated. A cluster is called an
active cluster as long as it remains isolated from any other cluster
or nucleation center at least by a layer of empty nearest
neighbors. Each cluster (active or dead) are marked with a unique
label. At the end of a MC step, if an active cluster is found
separated by a nearest neighbor bond from another cluster (active or
dead), they are merged to a single cluster and they are marked as a
single dead cluster. The growth of a dead cluster is seized for ever
as in GCM. If a peripheral site is rejected during the growth of an
active cluster, it will be not available for occupation by any other
growing cluster as in ordinary percolation (OP). However, such a site
can be occupied by a new nucleation center added externally. The
growth of a cluster stops due to the fact that either it becomes a
dead cluster by merging with another cluster or all its peripheral
sites become forbidden sites for occupation. The process of lattice
filling stops when no lattice site is available to add a nucleation
center. Time in this model is equal to the number of nucleation
centers added. Therefore, for any value of $g$, there will always be a
PT in the long time limit.

The model with $g=0$ naturally corresponds to the Hoshen-Kopelman
percolation as the instantaneous area fraction $p(t)$ reaches the OP
threshold $p_c(\rm OP) \approx 0.5927$ and exhibits a continuous
second order PT. For $g=0$, the area fraction $p(t)$ at time $t$ is
nothing but the number of nucleation centers added per lattice site up
to time $t$ whereas for $g\ne 0$, it is the number of occupied sites
per lattice site at time $t$. Such a continuous transition is expected
to occur as long as the growth probability $g$ remains below $p_c(\rm
OP)$. It can be noted here that in SFM, PT occurs only at unit area
fraction in the limit of growth probability tending to $0$. As $g$ is
increased beyond $p_c(\rm OP)$, large clusters appear due to the
merging of compact finite clusters originated from continuously added
nucleation centers. As a result, the system will lack small clusters
as well as power law distribution of cluster size at the time of
PT. Such a transition will occur with a discrete jump in the size of
the spanning cluster due to the merging of compact large but finite
clusters. Hence, it is expected to be a discontinuous first-order
transition. A smooth crossover from continuous transitions to
discontinuous transitions is then expected to occur as the growth
parameter $g$ will be tuned from below $p_c(\rm OP)$ to above $p_c(\rm
OP)$.

\section{Results and discussion}
Extensive computer simulation of the above model is performed on $2$D
square lattice of size $L\times L$. The size $L$ of the lattice is
varied from $L=64$ to $1024$ in multiple of $2$. Clusters are grown
applying periodic boundary condition (PBC) in both the horizontal and
the vertical directions. All dynamical quantities are stored as
function time $t$, the MC step or the number of nucleation centers
added. Time evolution of the cluster properties are finally studied as
a function of the corresponding area fraction $p(t)$ instead of time
$t$ directly. Averages are made over $2\times 10^5$ to $5\times 10^3$
ensembles as the system size is varied from $L=64$ to $1024$.

\begin{figure}[t]
\centerline{\hfill\psfig{file=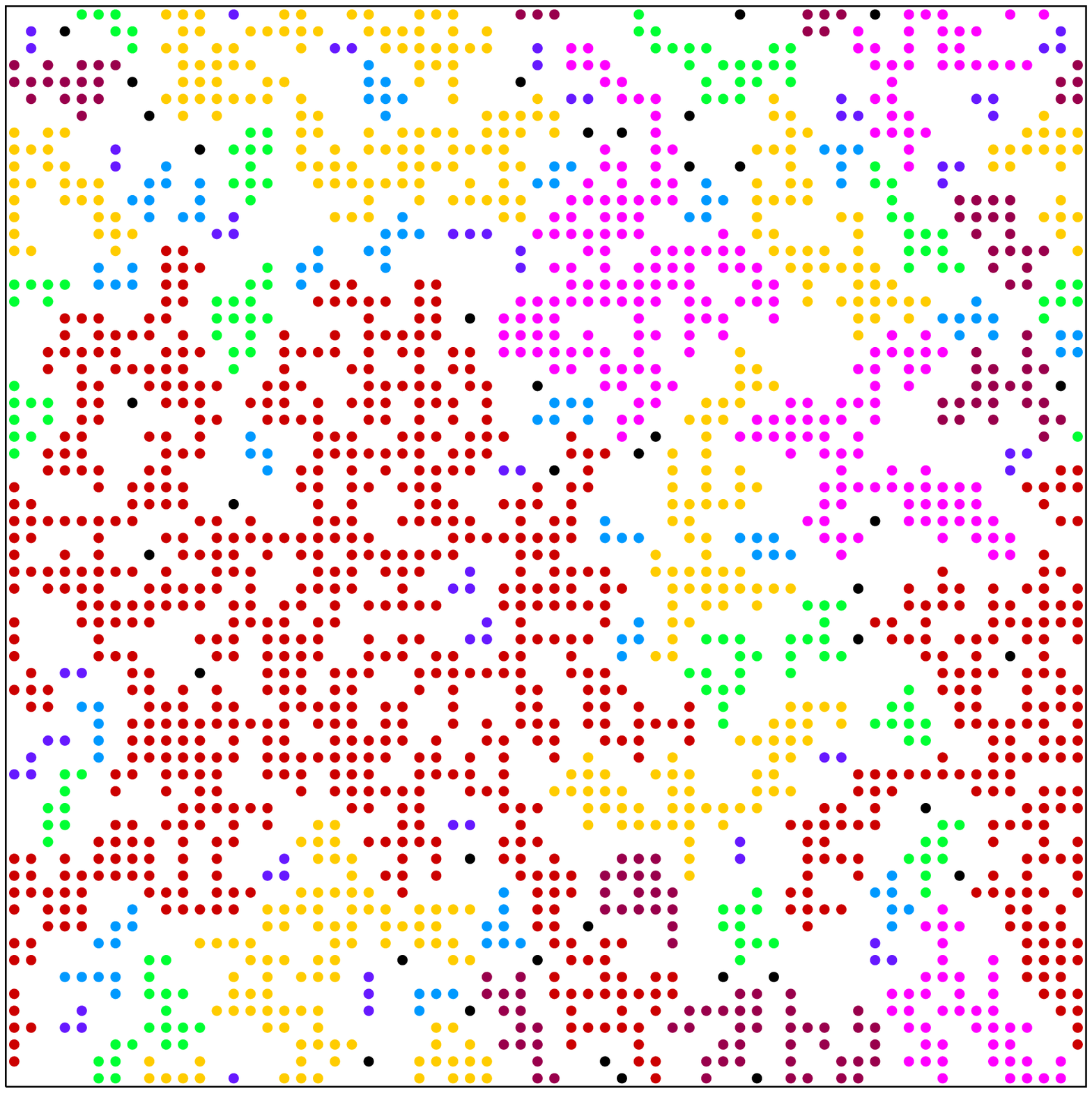,
    width=0.22\textwidth} \hfill\psfig{file=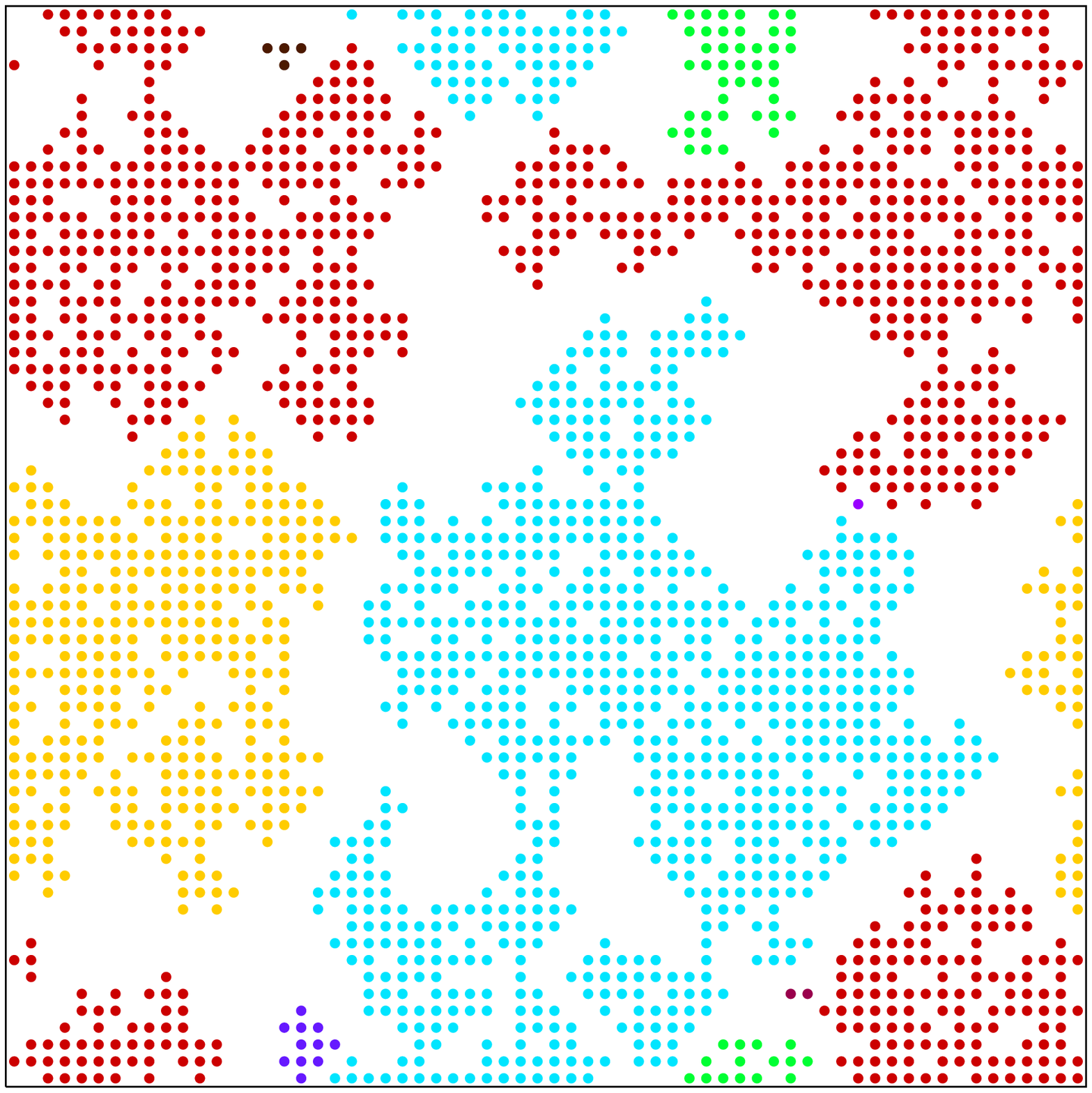,
    width=0.22\textwidth}\hfill} \centerline{\hfill (a) $g=0.4$,
  $t=859$ \hfill\hfill (b) $g=0.8$, $t=75$ \hfill}
\caption{\label{picl} (Color online) Snapshots of cluster
  configurations just before the appearance of a spanning cluster: (a)
  for $g=0.4$ at $t=859$ and (b) for $g=0.8$ at $t=75$ on a $2$D
  square lattice of size $L=64$. The red color shows the largest
  cluster and other different colors represent the presence of
  clusters of different sizes. Periodic boundary condition is applied
  in both horizontal and vertical directions during cluster growth.}
\end{figure}
 
\subsection{Cluster morphology and time evolution of the largest cluster}
The snapshots of cluster configurations are taken just prior to the
appearance of the spanning cluster in the system and are shown in
Fig.\ref{picl} for $g=0.4$ (a), and $g=0.8$ (b) on a $2$D square
lattice of size $L=64$. The largest cluster is shown in red and the
other smaller clusters of different sizes are depicted in other
different colors. At the lower growth probability $g=0.4$, clusters of
many different sizes exist along with a large finite cluster. Smaller
clusters are found to be enclaved inside the larger clusters. PT
occurs in the next step and no significant change in the largest
cluster size is expected as the largest cluster in the previous step
was about to span the lattice. Such continuous change in the largest
cluster size along with enclaved smaller clusters within it are
indications of continuous transition \cite{Sheinman2015}.  On the other hand,
as the growth probability is taken to be high $g=0.8$, clusters of
smaller sizes are merged with the fast growing other finite
clusters. As a result, only a few large compact clusters are found to
exist beside the newly planted nucleation centers. Clusters of
intermediate sizes are found to be absent. Enclave of smaller clusters
by the larger clusters has almost disappeared. As the clusters in cyan
and red are merged in the next step and generates a spanning cluster,
the PT occurs with a significant change in the size of the largest
cluster corresponding to a jump in the largest cluster size at the
time of transition. Appearance of compact large cluster with
discontinuous jump in the largest cluster size are indications of a
discontinuous transition \cite{bfwm031103}.

\begin{figure}[h]
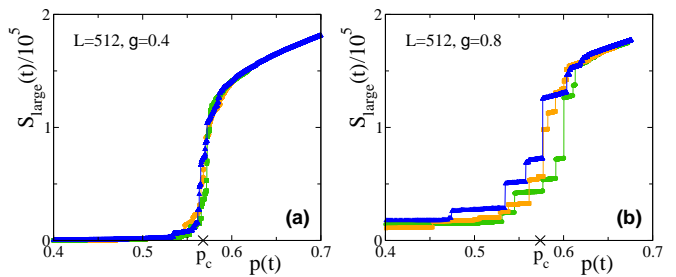

\centerline{\hfill\psfig{file=santra_fig2a.eps,width=0.24\textwidth}
\hfill\psfig{file=santra_fig2b.eps,width=0.24\textwidth}\hfill}
\caption{\label{smaxvt} (Color online) Time evolution of the largest
  cluster $S_{\rm large}(t)$ for a few samples are shown against the
  area fraction $p(t)$ for $g=0.4$ in (a) and for $g=0.8$ in (b) for a
  system of size $L=512$.}
\end{figure}

The time evolution of the size of the largest cluster $S_{\rm
  large}(t)$ is monitored against the instantaneous area fraction
$p(t)$ for three different samples for a given $g$. Their variations
are shown in Fig.\ref{smaxvt}(a) for $g=0.4$ and for $g=0.8$ in
Fig.\ref{smaxvt}(b) for a system of size $L=512$. The average area
fractions corresponding to the thresholds at which PT occur in these
systems with given $g$ are marked by the crosses on the respective
$p(t)$-axis. Around the respective thresholds, the evolution of the
size of the largest clusters for $g=0.4$ and $g=0.8$ are drastically
different. For $g=0.4$, the size of the largest cluster in different
samples are found to increase almost continuously with the
instantaneous area fraction $p(t)$ around the threshold. This
indicates a continuous PT to occur. However, for $g=0.8$, $S_{\rm
  large}$ grows with discontinuous jumps at the threshold with the
largest jump of the order of $10^5\gg L$. The effect would be more
prominent with higher values of $g$. This is another indication of a
discontinuous PT. It is then intriguing to study the model with
varying the growth probability $g$ and characterize the nature of
transitions at different regimes of $g$.

\subsection{Fluctuation in order parameter}
The dynamical order parameter $P_\infty(t)$, the probability to find a
lattice site in the spanning cluster, is defined as
\begin{equation}
\label{pinfty}
P_{\infty}(t)=\frac{S_{\rm max}(t)}{L^2}
\end{equation}
where $S_{\rm max}(t)$ is the size of the spanning cluster at time
$t$. The finite size scaling (FSS) form of $P_{\infty}(t)$ is given by
\begin{equation}
\label{fsspf}
P_{\infty}(t)=L^{-\beta/\nu} \widetilde{P}_{\infty}[\{p(t)-p_{c}(g)\}L^{1/\nu}]
\end{equation} 
where $\widetilde{P}_{\infty}$ is a scaling function, $\beta$ is the
order parameter exponent, $\nu$ is the correlation length exponent and
$p_{c}(g)$ is the critical area fraction for a given growth
probability $g$ at which a spanning cluster connecting the opposite
sides of the lattice appears for the first time in the
system. Following the formalism of thermal critical phenomena
\cite{bruce1992} as well as several recent models of percolation
\cite{Grassberger2011}, the distribution of $P_\infty$ is taken as
\begin{equation}
\label{pinfd}
P[P_{\infty}(t)]=L^{\beta/\nu} \widetilde{P}[P_{\infty}(t)L^{\beta/\nu}]
\end{equation}
where $\widetilde{P}$ is a scaling function. With such a scaling form
of $P[P_\infty(t)]$, one could easily show that
\begin{equation}
\langle P^2_{\infty}(t)\rangle \sim L^{-2\beta/\nu} \ \ \ {\rm and}
\ \ \ \langle P_{\infty}(t)\rangle^2 \sim L^{-2\beta/\nu}.
\end{equation}
The fluctuation in $P_\infty(t)$ at an area fraction $p(t)$ is defined
as
\begin{equation}
\label{exinf}
\chi_{\infty}(t)=\frac{1}{L^2} \left[\langle S^2_{\rm
    max}(t)\rangle-{\langle S_{\rm max}(t)\rangle}^2\right].
\end{equation}
The FSS form of $\chi_{\infty}(t)$ is given by
\begin{equation}
\label{chiinffss}
\chi_{\infty}(t) = L^{\gamma/\nu}
\widetilde{\chi}[\{p(t)-p_{c}(g)\}L^{1/\nu}]
\end{equation}
where $\gamma/\nu =d-2\beta/\nu$ is the ratio of the average cluster
size exponent to the correlation length exponent $\nu$, $d$ is space
dimension and $\widetilde{\chi}$ is a scaling function.
\begin{figure}[t]
  \centerline{\psfig{file=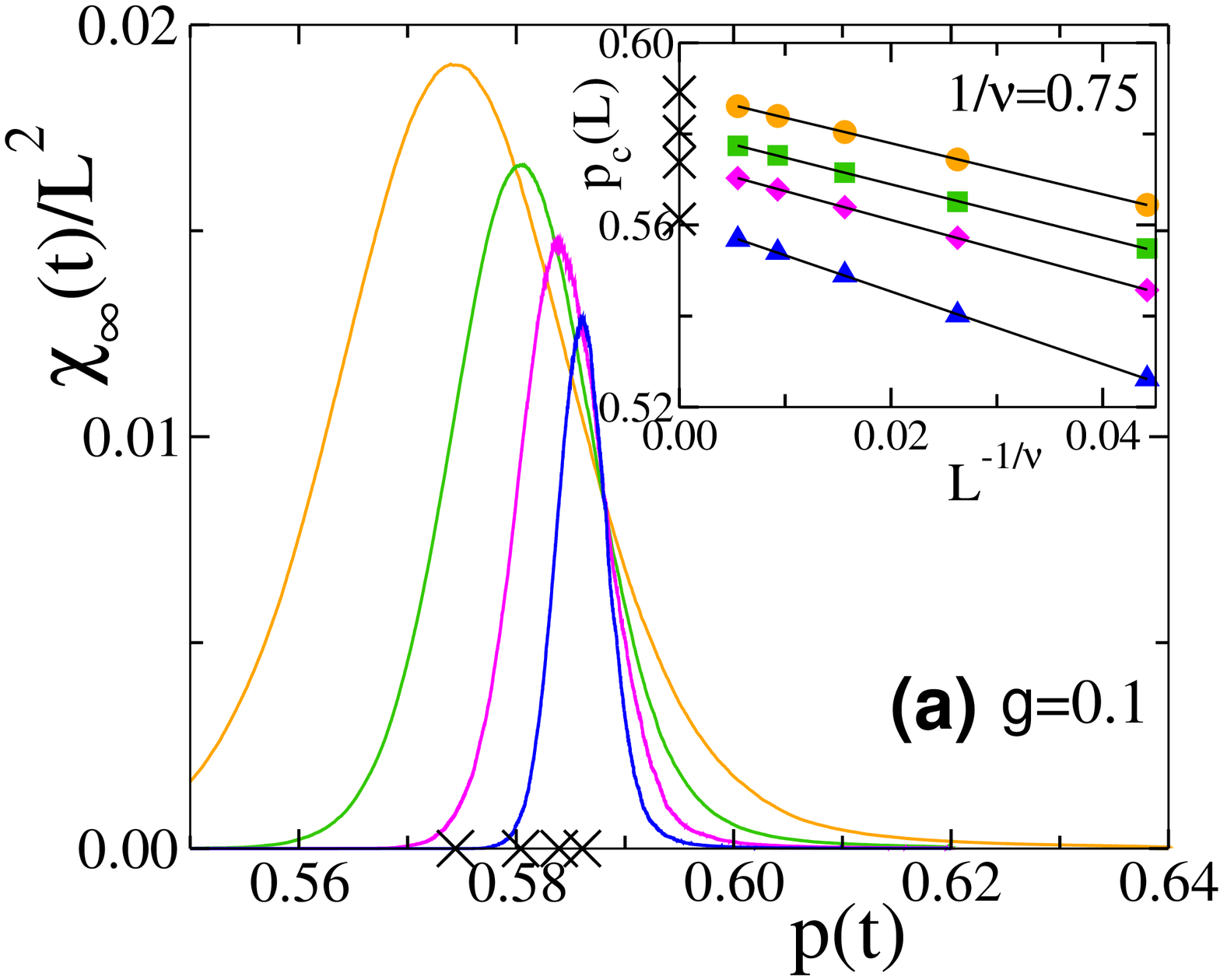,width=0.24\textwidth}
    \psfig{file=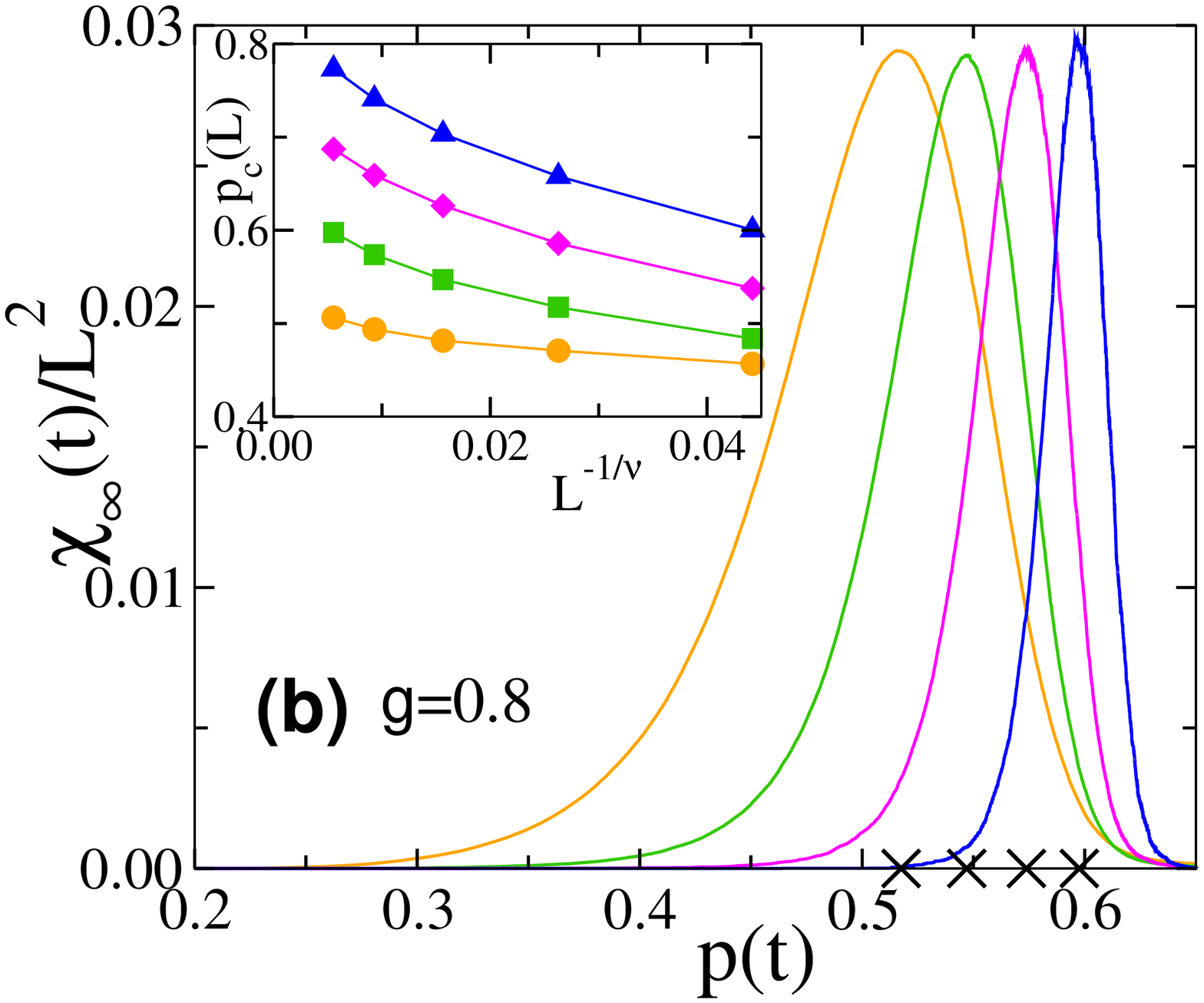,width=0.24\textwidth}}
  \caption{\label{xinf} (Color online) Plot of
    $\chi_{\infty}(t)/L^{2}$ against $p(t)$ for $g=0.1$ (a) and
    $g=0.8$ (b) for different lattice sizes $L$= $128$ (orange solid
    line), $256$ (green solid line), $512$ (magenta solid line) and
    $1024$ (blue solid line). Crosses on the $p(t)$-axis represent the
    thresholds $p_c(L)$. $p_c(L)$ is plotted against $L^{-1/\nu}$
    taking $1/\nu=0.75$, for $g=0.1(\CIRCLE)$, $0.3(\blacksquare)$,
    $0.4(\blacklozenge)$, $0.5(\blacktriangle)$ in the inset of (a)
    and for $g=0.7(\CIRCLE)$, $0.8(\blacksquare)$,
    $0.9(\blacklozenge)$, $1.0(\blacktriangle)$ in the inset of (b). }
\end{figure}
In Fig. \ref{xinf}, $\chi_{\infty}(t)/L^{2}$ is plotted against $p(t)$
for different lattice sizes at two extreme values of $g$: $g=0.1$ in
Fig.\ref{xinf}(a) and $g=0.8$ in Fig.\ref{xinf}(b). There are two
important features to note. First, each plot has a maximum at a
certain value of $p(t)$ for a given $g$ and $L$. The locations of the
peaks correspond to the critical thresholds $p_c(L)$ (marked by
crosses on the $p(t)$ axis) at which a spanning cluster appears for
the first time in the system. The critical area fraction $p_c(L)$ is
expected to scale with the system size $L$ as
\begin{equation}
\label{pcg}
p_c(L) - p_c(g) \approx L^{-1/\nu}
\end{equation}
where $\nu$ is the correlation length exponent, as it happens in OP
\cite{stauffer}. In the limit $L\rightarrow\infty$, the value of
$p_c(L)$ becomes $p_c(g)$, the percolation threshold of the model for
a given $g$. In the insets of respective figures, $p_c(L)$ is plotted
against $L^{-1/\nu}$ taking $1/\nu=0.75$, that of the OP, for
different values of $g$. The scaling form given in Eq.\ref{pcg} is
found to be well satisfied for $g\le 0.5$ with $1/\nu=0.75$, inset of
Fig.\ref{xinf}(a). For $g\le 0.5$, the linear extrapolation of the
plots of different $g$ intersect the $p_c(L)$ axis at different
$p_c(g)$ values. Whereas for $g\ge 0.8$, the data do not obey
Eq.\ref{pcg} and no definite $p_c(g)$ is found to exist in the
$L\to\infty$ limit. Such deviation from the scaling form given in
Eq.\ref{pcg} is found to occur for the systems those are grown with
$g>0.5$ too.

\begin{figure}[ht]
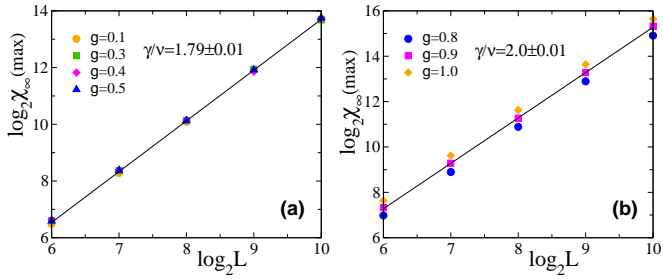

  \centerline{\psfig{file=santra_fig4a.eps,width=0.24\textwidth}
    \hfill\psfig{file=santra_fig4b.eps,width=0.24\textwidth}}
  \caption{\label{xmax} (Color online) Plot of $\chi_{\infty}(\rm
    max)$ against $L$ for $g\le 0.5$ is given in (a) and for $g\ge0.8$
    is given in (b). The straight lines of slope $\gamma/\nu=1.79$ and
    $\gamma/\nu=2.0$ in (a) and (b) respectively are guide to eye.}
\end{figure}
Second, the peak values of $\chi_\infty(t)/L^2$ are decreasing with
increasing $L$ for $g=0.1$ as in continuous transitions whereas they
remain constant with $L$ for $g=0.8$ as in discontinuous transitions
\cite{PhysRevLett.105.035701}. In order to extract the exponent
$\gamma/\nu$ for a system with a given value of $g$, the peak values
of the fluctuation $\chi_{\infty}(\rm max)$ are plotted against $L$
for $g\le 0.5$ in Fig.\ref{xmax}(a) and for $g\ge 0.8$ in
Fig.\ref{xmax}(b) in double logarithmic scale. The magnitudes of
$\chi_{\infty}(\rm max)$ are found to be independent of $g$ at a given
$L$ for $g\le 0.5$ whereas they increase with $g$ at a given $L$ for
$g\ge 0.8$. As per the scaling relation Eq.\ref{chiinffss}, a power
law scaling $\chi_{\infty}(\rm max)\sim L^{\gamma/\nu}$ is expected to
follow at the threshold. The exponent $\gamma/\nu$ is determined by
linear least square fit through the data points. For $g\le 0.5$, it is
found to be $\gamma/\nu=1.79 \pm 0.01$ whereas for $g\ge 0.8$, it is
found to be $\gamma/\nu=2.0 \pm 0.01$. The solid straight lines with
desire slopes in Fig.\ref{xmax}(a) and (b) are guide to eye. It is
important to note that the value of $\gamma/\nu$ for $g\le 0.5$ is
that of the OP ($43/24$) which indicates continuous transitions
whereas for $g\ge0.8$ it is that of the space dimension which
indicates discontinuous transitions. For $0.5<g<0.8$, the exponent
$\gamma/\nu$ is found to change continuously from $1.79$ to $2.0$
indicating a region of crossover. The values of $\gamma/\nu$ for
different values of $g$ are also verified by estimating the average
cluster size of all the finite clusters (excluding the spanning
cluster) at their respective percolation thresholds. However, there
are evidences in some other percolation models such as $k$-core
percolation \cite{PhysRevE.94.062307} that the scaling behavior of
order parameter fluctuation and that of the average cluster size are
not identical.

\begin{figure}[h]
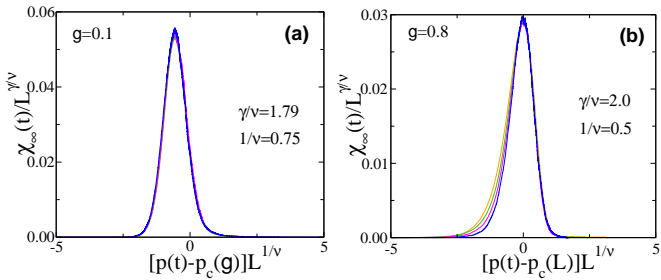

  \centerline{\psfig{file=santra_fig5a.eps,width=0.24\textwidth}
  \psfig{file=santra_fig5b.eps,width=0.24\textwidth}}
  \caption{\label{chiscl} (Color online) Plot of
    $\chi_{\infty}(t)/L^{\gamma/\nu}$ against the scaled variable
    $[{p(t)-p_c}L^{1/\nu}]$ for $g=0.1$ (a) and for $g=0.8$ (b)
    respectively.}
\end{figure}
The FSS form of $\chi_\infty(t)$ is verified plotting the scaled
fluctuation $\chi_\infty(t)/L^{\gamma/\nu}$ against the scaled
variable $[p(t)-p_c(g)]L^{1/\nu}$ for $g=0.1$ in
Fig.\ref{chiscl}(a). A good collapse of data is obtained for
$\gamma/\nu=1.79$ and $1/\nu=0.75$ as those of OP. Whereas, for
$g=0.8$, a partial collapse is obtained for the plots of
$\chi_\infty(t)/L^{\gamma/\nu}$ against the scaled variable
$[p(t)-p_c(L)]L^{1/\nu}$, as no $p_c(g)$ is available, taking
$\gamma/\nu=2.0$ and tuning the value of $1/\nu$ to $0.5$ as shown in
Fig.\ref{chiscl}(b). The collapse of the peak values confirms the
values of the scaling exponent $\gamma/\nu$ as $1.79$ for $g\le0.5$
and $2.0$ for $g\ge0.8$. Following the scaling relation
$\gamma/\nu=d-2\beta/\nu$, the exponent $\beta/\nu$ should be $0.105$
as that of OP for $g\le0.5$ and zero as that of a discontinuous PT for
$g\ge0.8$.

\begin{figure}[t]
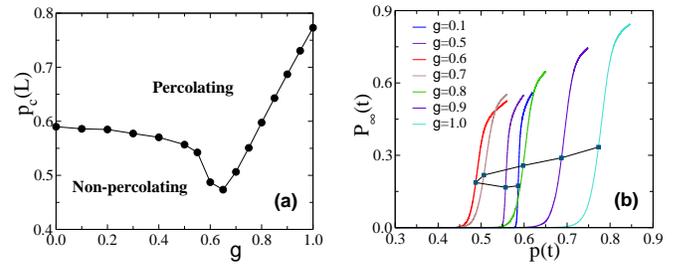

  \centerline{\psfig{file=santra_fig6a.eps,width=0.23\textwidth} \hfill
    \psfig{file=santra_fig6b.eps,width=0.23\textwidth}}
  \caption{\label{pcgPp} (Color online) (a) Plot of $p_c(L)$ against
    $g$.  (b) Plot of $P_\infty(t)$ against $p(t)$ for different
    values of $g$ for $L=1024$.}
\end{figure}
\subsection{Phase diagram}
A phase diagram separating the percolating and non-percolating regions
is obtained by plotting the variation of $p_c(L)$ against $g$ for a
system of size $L=1024$ in Fig.\ref{pcgPp}(a). It is interesting to
note that the critical area fraction has a minimum at a growth
probability little above the threshold of OP, $p_c$(OP)$\approx
0.5927$ and it is as low as $\approx 0.45$. It is obvious that area
fraction would be $\approx 0.6$ at the criticality when $g=0$. If $g$
is finite but small, growth of small clusters will stop mostly because
of less growth probability beside rarely merging with another small
cluster or a newly added nucleation center. A large number of smaller
clusters will be there in the system before transition and merging of
such small cluster will lead to a spanning cluster which will have
many voids in it. As a result, the area fraction will be less. Such an
effect will be more predominant when $g$ is around the percolation
threshold of OP as at this growth probability large fractal clusters
will be grown. PT occurs due to merging of such large fractal clusters
which will contain maximum void space in it. Hence, the area fraction
is expected to be the lowest. Beyond, such growth probability, compact
clusters start appearing which will occupy most of the space at the
time of transition. Area fraction will increase almost linearly with
$g$ in this regime. Such variation of $p_c$ is also observed in a
percolation model with repulsive or attractive rule in site occupation
\cite{refId0}.

The phase diagram is then complemented by the variations of
$P_\infty(t)$ against $p(t)$ for various values of $g$ which are shown
in Fig.\ref{pcgPp}(b). Not only the the critical threshold decreases
with increasing $g$ and takes a turn at $g \approx 0.65$ but also the
transitions become more and more sharper as $g$ increases beyond
$g\approx 0.65$. It is also interesting to note that values
$P_\infty(t)$ at $p_c$ also increases with increasing $g$ even when
the critical area fraction ($p_c$) is decreasing. Therefore the
spanning cluster mass is always increases with the growth probability
$g$.

\begin{figure}[t]
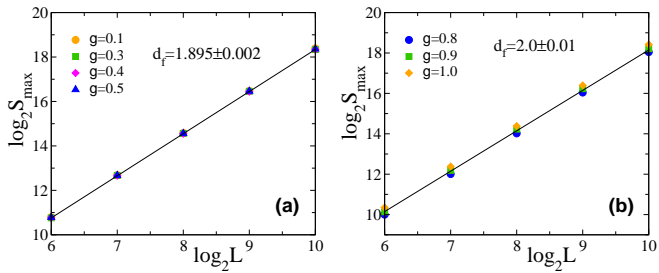

   \centerline{\psfig{file=santra_fig7a.eps,width=0.235\textwidth}
    \hfill\psfig{file=santra_fig7b.eps,width=0.235\textwidth}}
\caption{\label{smaxdf} (Color online) Plot of $S_{\rm max}$ against
  lattice sizes $L$ at their respective thresholds for $g\le 0.5$ in (a)
  and for $g\ge 0.8$ in (b). The solid straight line of slope $1.896$ in
  (a) and that of slope $2.0$ in (b) are guide to eye. }
\end{figure}

\subsection{Dimension of spanning cluster}
For system size $L\ll \xi$, the size of the spanning cluster
$S_{\rm max}$ at the criticality varies with the system size $L$ as
\begin{equation}
\label{fd}
S_{\rm max}\approx L^{d_f}
\end{equation}
where $d_f$ is the fractal dimension of the spanning cluster. Since
the clusters are grown here applying PBC, the horizontal and vertical
extensions of the largest cluster are stored. If either the horizontal
or the vertical extension of the largest cluster is found to be
greater than or equal to $L$, it is identified as a spanning
cluster. The value of $S_{\rm max}$ are noted at the respective
thresholds for several lattice sizes $L$ for a given $g$. For a
continuous PT, the spanning cluster is a random object with all
possible sizes of holes in it and is expected to be fractal whereas in
the case of a discontinuous transition it becomes a compact
cluster. The values of $S_{\rm max}$ are plotted against $L$ in double
logarithmic scale for the different values of $g\le 0.5$ in
Fig. \ref{smaxdf}(a) and for $g \ge 0.8$ in Fig. \ref{smaxdf}(b). For
$g\le 0.5$, $S_{\rm max}$ scales with $L$ as a power law with
$d_f=1.895\pm 0.002$ almost that of OP ($91/48$). On the other hand,
for $g \ge 0.8$, $S_{\rm max}$ scales with $L$ as a power law with
$d_f=2.0\pm 0.01$ as that of space dimension $d$. The solid lines with
desire slopes $1.896$ and $2.0$ in Fig.\ref{smaxdf}(a) and (b)
respectively are guide to eye. Thus for $g\le 0.5$, the spanning
cluster is found to be fractal as in OP whereas for $g\ge0.8$ they
appear to be compact as expected in a discontinuous transition. As a
result, there would be enclaves in spanning clusters for $g\le 0.5$
whereas such enclaves would be absent in the spanning clusters for
$g\ge 0.8$ as it is evident in the cluster morphology shown in
Fig. \ref{picl}. Such presence or absence of enclaves in the spanning
cluster determines whether it would be fractal or compact which
essentially determines the nature of the transition as continuous or
discontinuous \cite{Sheinman2015,PhysRevLett.106.095703}. In the
regime $0.5<g<0.8$ the dimension of spanning cluster $d_f$ changes
continuously from $d_f=1.895$ to $d_f=d=2.0$.

\begin{figure}[t]
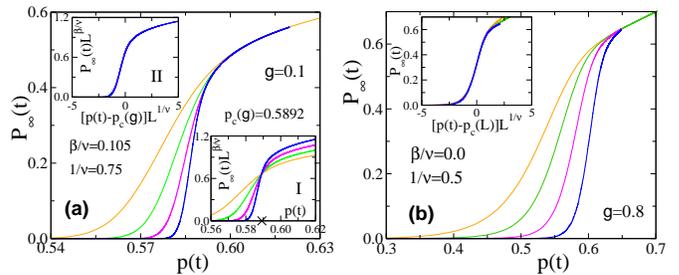

  \centerline{\psfig{file=santra_fig8a.eps,width=0.24\textwidth}
  \psfig{file=santra_fig8b.eps,width=0.24\textwidth}}
\caption{\label{pinf} (Color online) Plot of $P_{\infty}(t)$ against
  $p(t)$ for $g=0.1$ (a) and $g=0.8$ (b). In the inset-I of (a) for
  $g=0.1$, $P_{\infty}(t)L^{\beta/\nu}$ is plotted against $p(t)$
  taking $\beta/\nu=0.105$ and in the inset-II of (a) it is plotted
  against the scaled variable $[p(t)-p_c(g)]L^{1/\nu}$ taking
  $1/\nu=0.75$. In the inset of (b) for $g=0.8$,
  $P_{\infty}(t)L^{\beta/\nu}$ is plotted against
  $[p(t)-p_c(L)]L^{1/\nu}$ taking $\beta/\nu=0$ and $1/\nu=0.5$. Same
  set of color symbols of Fig. \ref{xinf} for different system sizes
  are used}
\end{figure}
\subsection{FSS of $P_\infty(t)$}
\begin{figure}[t]
\centerline{\psfig{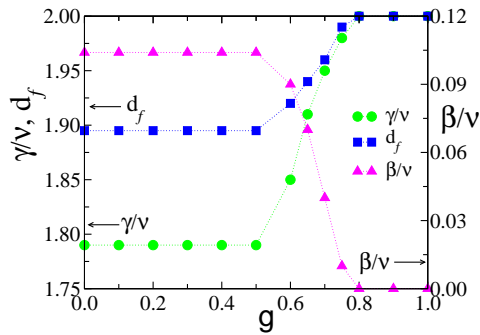}}
\caption{\label{crexp} (Color online) Values of the critical exponents
  against the growth probability $g$ are shown for $\gamma/\nu$ (green
  circle), $\beta/\nu$ (magenta triangle), and fractal dimension $d_f$
  (blue square). }
\end{figure}

The FSS form of $P_{\infty}(t)$ given in Eq.\ref{fsspf} as
$L^{-\beta/\nu}\widetilde{P}_{\infty}[\{p(t)-p_{c}(g)\}L^{1/\nu}]$
should scales with the system size $L$ as $P_{\infty}(t)\sim
L^{-\beta/\nu}$ at the criticality where $\beta/\nu = d-d_f$. As the
value of $d_f$ is found to be $1.895$ for $g\le 0.5$ and $2.0$ for
$g\ge0.8$, it is expected that the order parameter exponent
$\beta/\nu$ should be $0.105$ as that OP ($5/48$) for $g\le 0.5$ and
zero for $g\ge 0.8$ leading to discontinuous jump. A continuous
variation in $\beta/\nu$ is expected in the regime
$0.5<g<0.8$. Variation of $P_{\infty}(t)$ is plotted against $p(t)$
for different lattice sizes for $g=0.1$ in Fig. \ref{pinf}(a) and for
$g=0.8$ in Fig. \ref{pinf}(b). As the system size $L$ increases,
$P_{\infty}(t)$ becomes sharper and sharper for both $g=0.1$ and
$g=0.8$. However, the plots of $P_{\infty}(t)L^{\beta/\nu}$ cross at a
particular value of $p(t)$ corresponding to the critical threshold
$p_c(g)$ taking $\beta/\nu=0.105$ for $g=0.1$ as shown in inset-I of
Fig. \ref{pinf}(a). As $\beta/\nu=0$ for $g=0.8$, by no means they
could make cross at a definite $p(t)$. However, for $g=0.1$, after
re-scaling the $P_{\infty}(t)$ axis if the $p(t)$ axis is re-scaled as
$[p(t)-p_c(g)]L^{1/\nu}$ taking $1/\nu=0.75$ a complete collapse of
data occurs as shown in inset-II of Fig. \ref{pinf}(a). Whereas, for
$g=0.8$, collapse of $P_{\infty}(t)$ plots are obtained by re-scaling
only the $p(t)$ axis as $[p(t)-p_c(L)]L^{1/\nu}$ taking $1/\nu=0.5$ as
shown in the inset of Fig. \ref{pinf}(b). Such collapse of data not
only confirms the validity of the scaling forms assumed but also
confirms the values of the scaling exponents obtained. The
observations at $g=0.1$ are found to be the same for all $g\le 0.5$
and those are at $g=0.8$ are same for $g\ge 0.8$. Though discontinuous
jump in the order parameter is also observed in SFM, the PT is
characterized as continuous \cite{Chakraborty2014}. On the other hand,
in GCM, discontinuous transition is found to occur only in the
vanishingly small fixed initial seed concentration \cite{Tsakiris2011}
but for intermediate seed concentrations the transitions are found to
continuous that belong to OP universality class
\cite{argtouchstop,PhysRevE.87.022115}. For $0.5<g<0.8$, collapse of
data is observed for continuously varied exponents that depend on $g$
as also seen in Ref.\cite{PhysRevE.88.042122}. The variations of the
critical exponents $\gamma/\nu$, $\beta/\nu$ and fractal dimension
$d_f$ with the growth probability $g$ are presented in
Fig. \ref{crexp}. The values of the critical exponents clearly
distinguishes the discontinuous transitions for $g\ge0.8$ from the
continuous transitions for $g\le0.5$.

\subsection{Binder cumulant}
\begin{figure}[t]
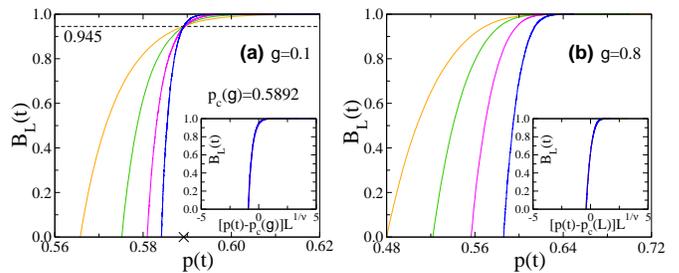

  \centerline{\psfig{file=santra_fig10a.eps,width=0.24\textwidth}
    \psfig{file=santra_fig10b.eps,width=0.24\textwidth}}
\caption{\label{binder} (Color online) Plot of Binder cumulant
  $B_L(t)$ against $p(t)$ for the different lattice sizes in (a) for
  $g=0.1$ and in (b) for $g=0.8$. The same color symbols of
  Fig. \ref{xinf} for different system sizes are used. Plot of
  $B_L(t)$ against the scaled variable $[p(t)-p_c]L^{1/\nu}$ are given
  in the respective insets taking appropriate values of $1/\nu$.}
\end{figure}

The evidences presented above indicate a continuous transition for
$g\le 0.5$ and a discontinuous transition for $g\ge0.8$. In order to
confirm the order of transition in different regimes of the growth
probability $g$, a dynamical Binder cumulant $B_L(t)$
\cite{PhysRevB.34.1841,1stfssbinder}, the fourth moment of $S_{\rm
  max}(t)$, is studied as function of area fraction $p(t)$. The
dynamical Binder cumulant $B_L(t)$ is defined as
\begin{equation}
\label{ebinder}
B_L(t)=\frac{3}{2}\bigg[1-\frac{\langle S_{\rm max}^4(t)\rangle}{3{\langle
      S_{\rm max}^2(t)\rangle}^2}\bigg].
\end{equation}
The cumulants when plotted against the area fraction $p(t)$ for
different system sizes $L$ are expected to cross each other at a
definite $p(t)$ corresponding to the critical threshold of the system
for a continuous transition whereas no such crossing is expected to
occur in the case of a discontinuous transition
\cite{Roy010101}. Though the cumulant has some unusual behavior
\cite{TSAI1998,botet}, it is rarely used in the study of recent models
of percolation. The values of $B_L(t)$ are plotted against $p(t)$ for
different system sizes $L$ in Fig. \ref{binder}(a) for $g=0.1$ and in
Fig. \ref{binder}(b) for $g=0.8$. For $g=0.1$, the plots of $B_L(t)$
cross at a particular $p(t)$ corresponding to $p_c(g)$, marked by a
cross on the $p(t)$-axis whereas for $g= 0.8$ no such crossing of
$B_L(t)$ is observed for different values of $L$. The value of the
Binder cumulant at the critical threshold $B_L(p_c)$ is found to be
$0.945$ as shown by a dotted line in Fig.\ref{binder}(a) for $g=0.1$
and remains close to this for other values of $g\le 0.5$. It is
verified that the value of $B_L(p_c)$ is same as that of ordinary site
percolation though it reported little less for the bond percolation
\cite{PhysRevE.71.016120}.

The FSS form of $B_L(t)$ is given by
\begin{equation}
\label{fssbc}
B_L(t)=\widetilde{B}[(p(t)-p_c)L^{1/\nu}]
\end{equation}
where $\widetilde{B}$ is a universal scaling function. The FSS form is
verified by obtaining a collapse of the plots of $B_L(t)$ against the
scaled variable $[p(t)-p_c(g)]L^{1/\nu}$ taking $1/\nu=0.75$ for
$g=0.1$. For $g=0.8$, however, such a collapse is obtained when the
cumulants are plotted against $[p(t)-p_c(L)]L^{1/\nu}$ taking
$1/\nu=0.5$. The data collapse is shown in the insets of the
respective plots. Such scaling behavior of $B_L(t)$ for $g=0.1$ is
found to occur for the whole range of $g\le 0.5$ and that of $g=0.8$
is found to occur for $g\ge0.8$. Once again, Binder cumulant provides
a strong evidence that the dynamical transition is continuous for
$g\le 0.5$ whereas it is discontinuous for $g\ge0.8$. For $0.5<g<0.8$,
a region of crossover, the cumulants do not cross at a particular
value of $p(t)$ rather they cross each other over a range of $p(t)$
values but do collapse when plotted against the scaled variable
$[p(t)-p_c(L)]L^{1/\nu}$ for the respective value of $1/\nu$ for a
given value of $g$. 

\subsection{Cluster size and order parameter distributions}
\begin{figure}[h]
  \centerline{\psfig{file=santra_fig11a.eps,width=0.235\textwidth}
    \psfig{file=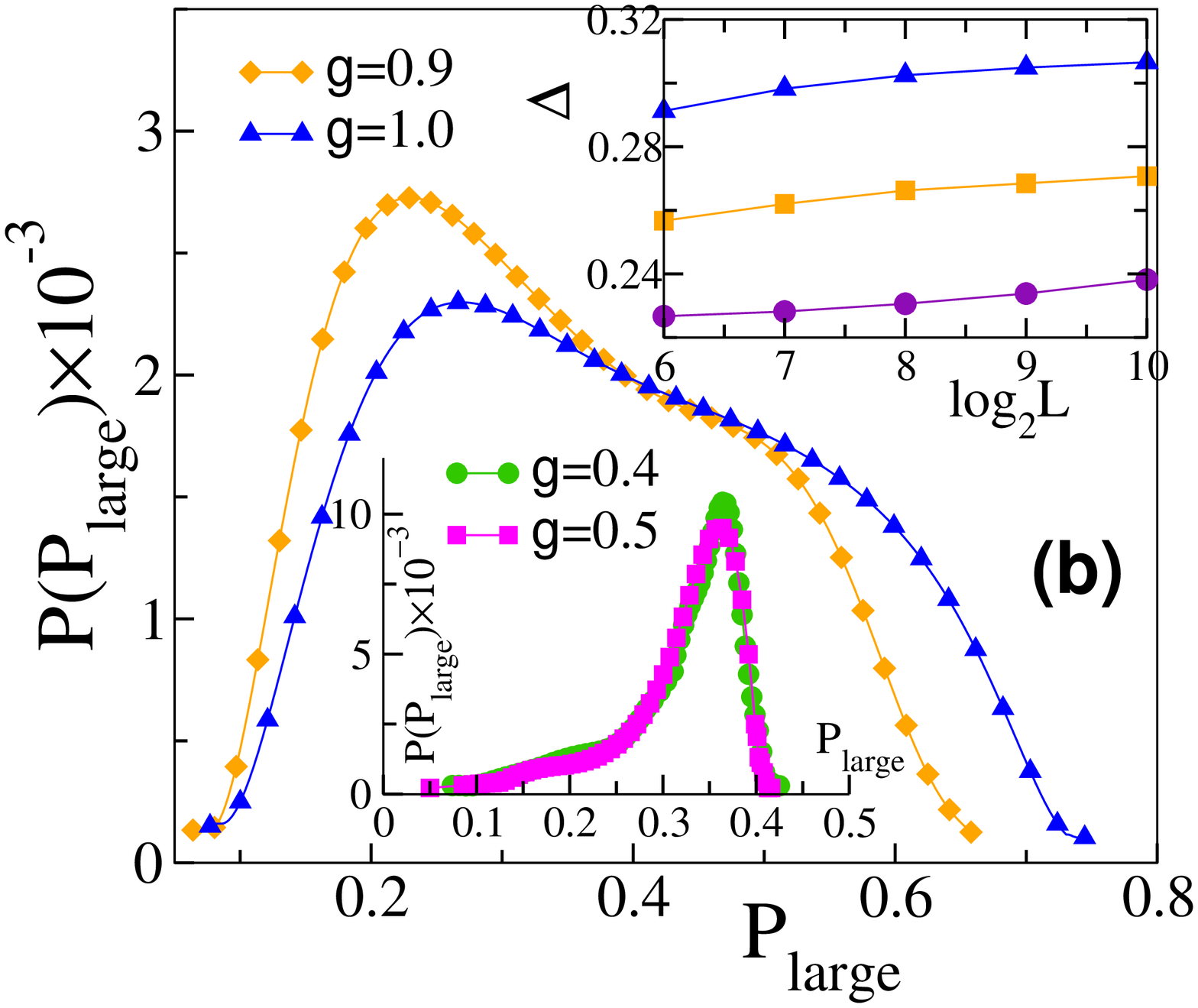,width=0.235\textwidth}}
\caption{\label{dis} (Color online) (a) Plot of $n_s(p_c)$ against $s$
  for different values of $g= 0.4[\CIRCLE]$ (green),
  $0.5[\blacksquare]$ (magenta), $0.9[\blacklozenge]$ (orange) and
  $1.0[\blacktriangle]$ (blue) for $L=1024$. Variation of the local
  slope $\tau_s$ $vs$ $s$ for the same set of $g$ values are shown in
  the inset of (a). (b) Plot of $P(P_{\rm large})$ against $P_{\rm
    large}$ for the corresponding systems in (a). For $g=0.4$ and
  $0.5$, plots are given as an inner plot using different scale. In
  the inset at top right corner of (b), plot of $\Delta$ against $L$
  for $g=0.8[\CIRCLE]$ (violet), $0.9[\blacksquare]$ (orange) and
  $1.0[\blacktriangle]$ (blue).}
\end{figure}

Power law distribution of cluster sizes at the critical threshold is
an essential criteria in a second-order continuous phase
transition. Following OP, a dynamical cluster size distribution
$n_s(t)$, the number of clusters of size $s$ per lattice site at time
$t$, is assumed to be
\begin{equation}
\label{nsg}
n_s(t)= s^{-\tau}{\sf f}[(p(t)-p_c)s^\sigma]
\end{equation}
where $\tau$ and $\sigma$ are two exponents and ${\sf f}$ is a
universal scaling function. For OP, an equilibrium percolation model,
the exponents are $\tau=187/91$ and $\sigma=36/91$
\cite{stauffer}. The distribution at the percolation threshold
$n_s(p_c)$ is expected to scale as $\approx s^{-\tau}$. The cluster
size distributions $n_s(p_c)$ are determined taking $p_c(g)$ as
threshold for $g\le0.5$ and taking $p_c(L)$ as threshold for $g\ge0.8$
for a system of size $L=1024$. The data obtained are binned of varying
widths and finally normalized by the respective bin widths. In
Fig. \ref{dis}(a), the distributions $n_s(p_c)$ are plotted against
$s$ in double logarithmic scale for $g\le 0.5$ ($0.4$ (green) and
$0.5$ (magenta)) and for $g\ge 0.8$ ($0.9$ (orange) and $1.0$ (blue))
for $L=1024$. It is clearly evident that the distributions for $g\le
0.5$ describes a power law behavior whereas for $g\ge0.8$ the
distributions develop curvature and deviate from power law scaling. In
the inset, the measured exponent $\tau_s=\partial
\log_{10}n_s(p_c)/\partial \log_{10}s$ is plotted against
$\log_{10}s$. The value of $\tau_s$ remains constant to $\approx
2.055$ as that of OP over a wide range of $s$ for $g\le0.5$ whereas
$\tau_s$ varies with $s$ for $g\ge0.8$ indicating no definite value of
$\tau$. The existence of a crossover from continuous transition of OP
type to a discontinuous percolation transition is further confirmed by
the value of $\tau$ in the different regimes of the growth probability
$g$. This is in contrary to the observations in SFM
\cite{Chakraborty2014} or cluster merging model \cite{Cho2016} where a
power law distribution of clusters size is found to occur beside
discontinuous transition.

Beside the cluster size distribution, distribution of order parameter
is also studied for different values of $g$ as usually it is studied
in thermal phase transitions \cite{bruce1992} where a bimodal
distribution of order parameter is expected in a discontinuous
transition whereas single peaked distribution is obtained in a
continuous transition. An ensemble of largest clusters on different
configurations are collected at the percolation threshold of a given
$g$ and the values of the order parameter $P_{\rm large}=S_{\rm
  large}/L^2$ are estimated. A probability distribution $P(P_{\rm
  large})$ is then defined as
\begin{equation}
\label{fssop}
P(P_{\rm large})\sim L^{\beta/\nu}\widetilde{P}[P_{\rm large}L^{\beta/\nu}]
\end{equation}
where $\widetilde{P}$ is a scaling function. Bimodal nature of
$\widetilde{P}$ is found to be a powerful tool to distinguish
discontinuous transitions from continuous transitions in some of the
recent percolation models
\cite{Grassberger2011,Manna20122833,Tian2012286,Roy010101}. The
distributions of $P(P_{\rm large})$s are plotted in Fig. \ref{dis}(b)
for different values of $g$. For $g\ge 0.8$, instead of sharp bimodal
distributions, broad distributions with two weak peaks are
obtained. No FSS of the distributions is found as given Eq.\ref{fssop}
but the width of the distribution $\Delta=2[\langle P_{\rm
    large}^2\rangle-\langle P_{\rm large}\rangle^2]^{1/2}$ for a given
$g$ is found to increase with the system size $L$, shown in the inset
of Fig. \ref{dis}(b), as a signature of discontinuous transition. For
a given $L$, the width of the distributions $\Delta$ is also found to
increase with $g$. However, the distributions $P(P_{\rm large})$ for
$g\le0.5$ are found to be single humped and follow the scaling form
given in Eq.\ref{fssop} as shown in the other inset. The width of the
distributions for a given $g\le0.5$ is found to decrease with $L$. The
model, thus, exhibits characteristic properties of discontinuous
transition for $g\ge0.8$ and those of continuous transition for $g\le
0.5$.

\section{Conclusion}
In a dynamical model of percolation with random growth of clusters
from continuously implanted nucleation centers through out the growth
process with touch and stop rule, a crossover from continuous to
discontinuous PT is observed as the growth probability $g$ tuned from
$0$ to $1$. For $g\le 0.5$, the order parameter continuously goes to
zero and the geometrical quantities follow the usual FSS at the
critical threshold with the critical exponents that of OP. The cluster
size distribution is found to be scale free and a single humped
distribution of order parameter occurred in this regime of $g$. On the
other hand, for $g\ge0.8$, the PT occurs with a discontinuous jump at
the threshold, the order parameter fluctuation per lattice site
becomes independent of system size, the spanning cluster becomes
compact with fractal dimension $d_f=2$ as that of discontinuous
transitions. No scale free distribution is found for the cluster sizes
and the order parameter distribution is weakly double humped broad
distribution of increasing width with the system size. The order of
transitions in different regimes of $g$ are further confirmed by the
estimates of Binder cumulant. The intermediate regime of growth
probability $0.5<g<0.8$ remains a region of crossover without a
definite tricritical point.

\bibliographystyle{aip}
\bibliography{ref}

\end{document}